\documentclass{svmult}

\usepackage{makeidx}         % allows index generation
\usepackage{graphicx}        % standard LaTeX graphics tool
\usepackage{multicol}        % used for the two-column index
\usepackage[bottom]{footmisc}% places footnotes at page bottom
\makeindex
\begin{document}

\title*{Stability and dewetting of thin liquid films}
\author{Karin Jacobs,\inst{1}\ Ralf Seemann,\inst{1,2}\ and Stephan Herminghaus\inst{2}}
\institute{Dept. of Experimental Physics, Saarland University,
D-66123 Saarbr\"{u}cken, Germany. \and Max-Planck-Institute for
Dynamics and Self-Organization, D-37073 G\"{o}ttingen, Germany.}

\authorrunning{Karin Jacobs, Ralf Seemann, Stephan Herminghaus}

\maketitle

\begin{abstract}
The stability of thin liquid coatings is of fundamental interest in
everyday life. Homogeneous and non-volatile liquid coatings may
dewet either by heterogeneous nucleation, thermal nucleation, or
spinodal dewetting. Wetting and dewetting is explained on a
fundamental level, including a discussion of relevant interactions.
The chapter will also address the various dewetting scenarios and
explain how the effective interface potential governs the behavior
obtained for various stratified substrates and film thicknesses.
\end{abstract}

%\body

\section{Introduction}
'To wet or not to wet' \cite{Evans_Chan} is the question this chapter tries to answer.
 It concerns the fundamental aspects of the stability of thin liquid
 films or coatings. The liquid in question can be any homogeneous liquid\footnote{homogeneous on the length scale set by the thickness of the liquid film under consideration.
 In particular, demixing or segregation effects are thereby assumed to be negligible.}, such as, e.g.,
 an aqueous solution, an oil, or a polymer melt. Whether or not a liquid wets a given surface is not only of
 fundamental interest, but also of substantial technical importance. The authors have been approached by numerous
 companies with interesting problems concerning wettability, ranging from the automotive
 industry, the pharmaceutical and chemical industry, to the food processing, printing, and textile industry.
In principle, the general answer to all wettability problems is
simple: a liquid wets a surface if it can gain energy by enlarging
the interface with that substrate. To provide this answer for a
specific system under study, however, may be quite cumbersome. Below
we provide what may be seen as a map to reader to help with a first
encounter with thin liquid film stability problems. For the sake of
clarity, we restrict our considerations to liquids on solid,
homogeneous substrates.

\section{Experimental model systems for simple liquids}

In order to study the basic mechanisms of a certain class of
phenomena, it is of central importance to identify suitable model
systems. Since the main problems in the controlled preparation of
thin films and surfaces lies in their inherent propensity to become
contaminated by dust or other impurities, the first quantitative
studies of surface forces in wetting films have been performed with
cryogenic systems, in which impurities are naturally frozen out
\cite{SabiskyAnderson,RutledgeTaborek,HerminghausAnnalen}. However,
it has been realized that these systems involve strong exchange of
material between the liquid and vapor phase and thus are not well
suited for studies of dynamic aspects, such as dewetting and
structure formation phenomena.

Thin laser-annealed metal films were candidates,\cite{Bischof,Herminghaus98} yet the same time, techniques of preparing systems which involve more
complex liquids, such as polymer melts, in a well-defined and clean
way became progressively available. As a result, the numerous
benchmark studies of the basic mechanisms of dewetting have been
performed with polymer melts (see e.g. the review articles of D.G.
Bucknall\cite{Bucknall2004} and P. M\"uller-Buschbaum\cite{MuBu03}).
Systematic experimental studies were begun somewhen around the 1990s
\cite{Redon91,Reiter92,Rachel94,Xie98,Jacobs98} inspired by P.-G. de Gennes'
theoretical work.\cite{Gennes79,Gennes85} Polymer melts are on the
one hand close to application (coatings, photo resist), yet on the
other hand easily controllable in the experiments. Polymers such as
polystyrene (PS) are very suitable model liquids since they have a
very low vapor pressure in the melt, and mass conservation can
safely be assumed. Moreover, they are chemically inert, non-polar,
and their dynamics can be tailored by choosing different chain
lengths and annealing temperatures. For molecular weights below the
entanglement length ($\sim$17~kg/mol), the melt can safely be treated
as simple (Newtonian) liquid for the low shear rates in dewetting
experiments\cite{Seemann01,Renate2005}. Below the glass transition
temperature $T_g$, the films are glassy and can be stored for
subsequent analysis\footnote{It is important to note that $T_g$ of thin films can be  substantially different (mostly lower) than the bulk value.\cite{Keddie94,DeGennes00,Forrest01,Herminghaus01Eur,HerminghausLandfest,Baschnagel05,Alcoutlabi05}}. PS in atactic steric configuration moreover does
not exhibit any tendency for crystallization.

As substrates, typically hydrophobized Si wafers with their natural
amorphous Si dioxide layer (SiOx) have been widely used. the present
chapter thus concentrates on Si wafers as substrates, which can be
purchased with a very small surface roughness (rms roughness smaller
than 0.2~nm). The wettability of these substrates can be tuned by
chemically grafting monomolecular functional layers onto the native
silicon oxide overlayer. For instance, hydrophobization is often
achieved by grafting a self-assembled monolayer of
octadecyl-trichlorosilane\cite{Wassermann,Brzoska} onto the wafer.

For dewetting experiments it is necessary to prepare a thin liquid
film in a non-equilibrium state on a substrate. Usually, a thin
polymer coating is prepared from a solvent solution by one of the
following standard techniques: spin coating, dip coating, or
spraying\footnote{For all these techniques, the surface to be coated
must be wettable by the polymer solution. Otherwise, the film must
be prepared on another substrate, e.g. mica, and then transferred to
the surface of interest.}. The solvent
evaporates during the preparation procedure, leaving behind a
smooth, glassy polymer layer. By spin coating a solution of
polystyrene in toluene one can easily achieve a PS layer of
thickness in the range of a few nanometers up to several
micrometers. The roughness of the polystyrene layer is then similar
to the underlying substrate. For experimental details and caveats of
preparation, the reader is referred to the pertinent literature´.
\cite{Brochard87,Jacobs98,JacobsSchatz98,Reiter92,See2001,Seemann01,Seemann2001}
 The thickness of the films can be determined
by standard techniques, such as ellipsometry\cite{AzzamBashara}. The
dewetting process can be studied by optical microscopy and/or atomic
force microscopy (AFM). Most studies use AFM in non-contact mode
operation to avoid any damage of the soft surface, which even allows
to image the dewetting scenario in situ\cite{Becker03,Renate2007b}.

\begin{figure}[h!]
\begin{center}
\includegraphics [width=1.0\textwidth]{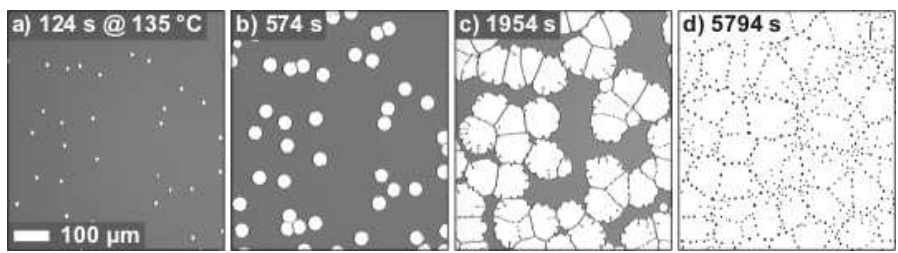}
\end{center}
\caption{\footnotesize Pictures series taken by a light microscope: a 80 nm
thick polystyrene film of 65 kg/mol molecular weight is dewetting at
135$^{\circ}$C from a hydrophobized silicon substrate for approximately (from left to right) 2 min, 10 min, 30 min, and 100 min.}\label{PS-Film}
\end{figure}

To induce dewetting, the films are heated above the glass transition
temperature. Figure~\ref{PS-Film} shows a series of optical
micrographs of an 80 nm thick PS film of molecular weight of 65
kg/mol (`PS(65k)') dewetting a silanized Si wafer. The series
depicts a pattern formation process that is typical for most
dewetting films, whether for a water film on a waxed surface, or a
coating on a dusty substrate.

The process of dewetting can be divided into three stages: in the
early stage, holes are generated by a rupture process, c.f.
Fig.~\ref{PS-Film}a; in the intermediate stage, the radius of the
holes increases, leading to hole coalescence, c.f.
Fig.~\ref{PS-Film}b-c. In the intermediate stage, the focus is on
the dynamics involved in the dewetting process, its impact on the
hole profiles and on its influence on dewetting patterns. From the
dynamics of hole growth
\cite{Brochard87,Brochard92,Redon94,Brochard94,Reiter96,JacobsSchatz98,Reiter00,ReiterNat2005,Renate2007} as well as from the shape of the
liquid rim surrounding the hole one can get information about the
slip or no-slip boundary condition of the liquid close to the solid
substrate \cite{Renate2005,Renate2006,Renate2007a}. In
the late stage, the straight ribbons that separate two coalescing
holes decay into droplets due to the Rayleigh-Plateau
instability\cite{Rayleigh78}, c.f. Fig.~\ref{PS-Film}c-d. Slight
differences in the size of the droplets cause slight pressure
differences, leading small droplets to shrink and large droplets to
grow (Ostwald ripening\cite{Ostwald}). Due to the small vapor
pressure and the low mobility of the polymer molecules, the ripening
process is extremely slow. Most experiments are stopped before the
'final state', i.e. one single drop on the surface, is reached. In
what follows, we focus on the initial stage of dewetting: why does
the polystyrene film dewet the hydrophobized Si wafer at all? The
next section will be dedicated to the discussion of the various
energy contributions involved, and their mutual balance.

\section{The energy balance}
In a Teflon$^{\footnotesize \copyright}$ coated frying pan a stable
oil layer can only be achieved if one pours enough oil into the pan.
Gravity then stabilizes the thick oil film. If the oil film thins
well below the so-called capillary length $\lambda_{cap}$, capillary
(intermolecular) forces dominate over gravitational forces and the
film may dewet \cite{HerminghausBrochard}. The capillary length is
given by $\lambda_{cap}$=$\sqrt{\sigma_{lv} /\rho g}$ (for olive
oil: $\lambda_{cap} \approx$ 1.7~mm, for water: $\lambda_{cap}
\approx$ 2.7~mm), where $\sigma_{lv}$ is the liquid/vapor surface
tension and $\rho$ the density of the liquid. Therefore, dewetting
of liquid films driven by intermolecular forces can only occur for
`thin' films, i.e. films thinner than $\lambda_{cap}$.

A liquid can be metastable and then needs a nucleus to induce dewetting.
In the following we will focus on liquid films thinner than $\lambda_{cap}$
and discuss under which conditions a thin liquid layer will be stable, metastable or
unstable on top of a substrate.

\subsection{The Young equation}

Dewetting is a dynamic process that begins in a non-equilibrium situation \cite{Young,Safran,Israelachvili,Schick,Dietrich,Gennes85,Oron97,Oron00}, namely the flat film on the surface and ends when reaching an equilibrium state, one droplet or a set of droplets \footnote{Equilibrium situation actually is one single droplet, yet it normally takes too long to reach this state. Before, a network of droplets is formed.}. Thus, let us first start with a droplet, c.f. Fig.~\ref{wetting}:

\begin{figure}[h!]
\begin{center}
\includegraphics[width=0.8\textwidth]{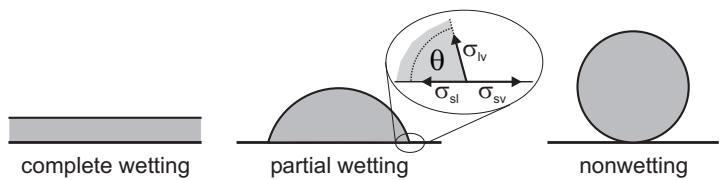}
\end{center}
\caption{\footnotesize Sketch of a liquid drop atop of a solid
substrate. Complete wetting is characterized by a contact angle
$\theta = 0$, partial wetting by $0 < \theta < \pi$, and
nonwetting by $\theta = \pi$ (from left to right)}\label{wetting}
\end{figure}

A droplet on a homogeneous surface usually exhibits the form of a spherical
cap and the tangent to the droplet at the three phase contact line includes
an angle $\theta$ with the substrate.  In equilibrium, the angle is given by
a balance of macroscopic forces:
\begin{equation} \label{Youngeq}
\cos \theta = \frac{\sigma_{sv} - \sigma_{sl}}{\sigma_{lv}}
\end{equation}
This is the Young equation \cite{Young} of 1805, where
$\sigma_{sv}$ and $\sigma_{sl}$ are the solid/vapor and solid/liquid
interfacial energies, and $\sigma_{lv}$ is the liquid/vapor
interfacial energy (or tension).

For $\theta= 0^\circ$, i.e. $\sigma_{sv} - \sigma_{sl} \geq
\sigma_{lv}$ the droplet will spread and will completely wet the
substrate. For $0^\circ < \theta < 180^\circ$ one speaks of partial
wetting, and for $\theta= 180^\circ$ of non-wetting. All three cases
are sketched in Fig.~\ref{wetting}. In case of water (resp. oil), a
surface is termed 'hydrophilic' ('oleophilic') if $0^\circ
\leq\theta < 90^\circ$ and 'hydrophobic' ('oleophobic') for
$90^\circ \leq\theta \leq 180^\circ$ \footnote{It should be noted
that non-wetting has never been observed on flat surfaces, although
some systems (like, e.g., mercury on glass) come quite close.}. In
other words, the contact angle in Young's equation is determined by
the free energies of interfaces between semi-infinite media.

\begin{figure}[h!]
\begin{center}
\includegraphics[width=0.8\textwidth]{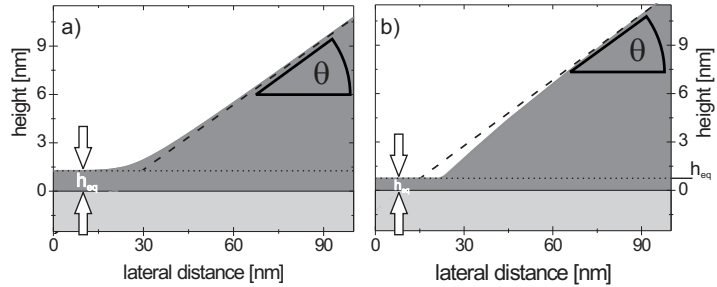}
\end{center}
\caption{\footnotesize Droplet profiles in the vicinity of the three
phase contact line as expected for a) an unstable and b) a
metastable situation. The profiles have been calculated for the case
of a) a $d_{SiOx}= 191~nm$ on Si and b) a $d_{SiOx}= 1.7~nm$ SiOx layer on Si,
the potentials of which are shown in Fig.~\ref{Lambda_Grenzpot}. The
dashed line marks the macroscopic contact angle $\theta=
7.5^{\circ}$ for a PS drop and the arrows mark the equilibrium
wetting layer thickness $h_{eq}$}\label{drop-profile}
\end{figure}

This is, however, not the full story. Although a first glance at
Fig.~\ref{wetting} suggests that the free energy of a homogeneous
film should be written as $\sigma_{\rm film} = \sigma_{sl} +
\sigma_{lv}$, this approach neglects possible interactions of the
two interfaces (solid/liquid and liquid/vapor) with each other,
across the liquid film. Such interactions may come up by virtue of
van der Waals forces between the involved materials, which may span
several tens of nanometers. In the vicinity of the three phase
contact line, the interfaces can thus deviate significantly from a
straight intersection at Young's angle (cf. Fig.~\ref{drop-profile}). We
will come back to this point later, after having discussed in some
detail the long-range forces of a stratified system.

These forces are of similar importance for predicting the
stability of a liquid film. Their dependence on film thickness
determines not only the linear stability against small perturbations
of the free surface, but also the scenario by which the initially
uniform film is transformed into its equilibrium state, which
consists of droplets of contact angle $\theta$ on the surface
\cite{droplets,Sharma98,Quere04}. For example, a droplet of photo resist on
a semiconductor may exhibit an equilibrium contact angle $\theta$ =
20$^{\circ}$ and therefore wets the surface only partially. What
does this mean for a photo resist film that was prepared by a
non-equilibrium technique (spin coating) on top of the
semiconductor? According to Young's equation, the film is not
stable. How will the film decay into droplets and at what speed will
this process take place at a given temperature? Will there be time
to dry or cure the photo resist before it dewets?

Depending on the thickness of the photo resist, the answers will be
different, though Young's contact angle $\theta$ is always the same!
Macroscopic and molecular terms describing stability conditions
often were mixed up and have led to confusion. In what follows, we
shall discuss which forces are to be considered, and what typical
length scales are involved.

\subsection{The effective interface potential}

Following the preceding section, we {\it define} the so-called
effective interface potential $\phi(h) \equiv \sigma_{sl} +
\sigma_{lv} - \sigma_{\rm film}$. It comprises both  short-range and
long-range interactions and is defined as the excess free energy per
unit area which is necessary to bring two interfaces from infinity
to a certain distance. From its very definition, it is clear that
$\phi(h) \rightarrow 0$ as $h$ tends to infinity. The excess free
energy of an infinitely thick film is thus given by the sum of the
free energies of its two interfaces, in accordance with intuition.

For dielectric systems, there are only two relevant types of
interactions, steric repulsion and van der Waals forces\footnote{A
detailed description for numerous different situations involving
e.g. polar molecules or hydrogen-bonds or the interactions of a
colloidal sphere interacting with a metallic surface can be found in
the textbook of J. Israelachvili\cite{Israelachvili}.}.
\begin{equation} \label{effective_interface}
\phi(h) = \phi(h)_{steric} + \phi(h)_{vdW}
\end{equation}

\begin{figure}[h!]
\begin{center}
\includegraphics[width=0.6\textwidth]{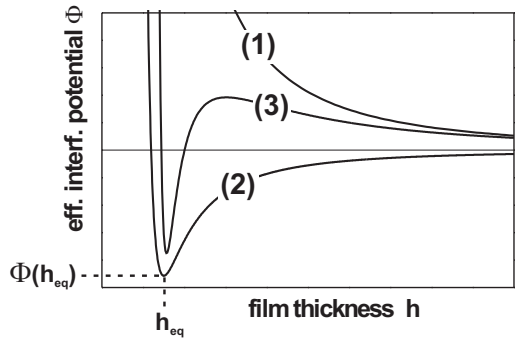}
\end{center}
\caption{\footnotesize Sketch of the effective interface potential
$\phi$ as a function of film thickness. Line (1) denotes the
stable case, line (2) the unstable one and curve (3) the
metastable case. The thickness of the stable wetting layer is termed
$h=h_{eq}$, and is typically in the order of some nm, the depth of
the global minimum of $\phi$ is named $\phi(h_{eq})$.}
\label{Grenzpot}
\end{figure}

In Fig.~\ref{Grenzpot} three typical curves of $\phi(h)$ are
sketched to illustrate the general principle. Line (1) characterizes
a film that is stable on the substrate, since energy would be
necessary to thin the film. The equilibrium film thickness is
infinite. The two other curves exhibit a global minimum of $\phi(h)$
at $h = h_{eq}$: Curve (2) characterizes a film that is unstable,
whereas line (3) describes a metastable film.

It is readily shown by a linear stability analysis
\cite{Vrij,Ruckenstein,Williams82} that if the second
derivative of $\phi(h)$ with respect to $h$ is negative ($\phi''(h)<
0$), unstable modes exist whose amplitudes grow exponentially
according to $\exp(t/\tau)$, where $\tau$ is the growth time that is
characteristic for the respective mode. Furthermore, there is a
characteristic wavelength $\lambda_{s}$  of these modes the
amplitude of which grows fastest and will therefore dominate the
emerging dewetting pattern.

The linear stability analysis also reveals that the spinodal
wavelength $\lambda_{s}$ is linked to the (second derivative of the)
effective interface potential \cite{Vrij,Ruckenstein}:
\begin{equation} \label{lambda}
\lambda_s (h) = \sqrt {\frac{- 8 \pi ^2 \sigma_{lv}}{\phi''(h)}} \; .
\end{equation}
The spinodal wavelength $\lambda_{s}$ is the key to clearly identify
a spinodal dewetting process in experimental systems.\footnote{The
name 'spinodal dewetting' has been coined due to the analogy to
spinodal decomposition of a blend of incompatible liquids, which
occurs if the second derivative of the free energy with respect to
the composition is negative\cite{Mitlin93}. As in dewetting, this
mechanism is clearly distinct from heterogeneous nucleation, in
which randomly distributed impurities dominate the emerging pattern.
\cite{Jacobs98}}

One question is now to be answered: What links the effective
interface potential to Young's contact angle? The effective
interface potential can describe a non-equilibrium situation, yet
Young's contact angle is only defined for the equilibrium. A.
Frumkin has published in 1938 that\cite{Frumkin}
\begin{equation} \label{Pottiefe}
\frac{\phi(h_{eq})}{\sigma_{lv}} = \cos \theta - 1 \; .
\end{equation}
What is the consequence of Eq.~(\ref{lambda}) and
Eq.~(\ref{Pottiefe})?  Determining the spinodal wavelength
$\lambda_s$ as a function of film thickness $h$ enables us to gain
insight into the course of $\phi''(h)$. By additionally measuring
the equilibrium layer thickness $h_{eq}$ and the contact angle
$\theta$, it is possible to reconstruct the complete effective
interface potential,\cite{See2001,Seemann2001,Herminghaus98} c.f.
Fig.~\ref{Lambda_Grenzpot}, as will be shown later.

\vspace{0.5cm}

The steric and the van der Waals part of the effective interface
potential are characterized by different exponents and different
interaction constants, therefore we will discuss the interactions
separately:

\textbf{Steric repulsion} and chemical interactions are relevant
only within a few \AA ngstroms of film thickness, and the resulting
force is therefore termed `short-range force'. The repulsion is due
to overlapping electron shells and is typically described by a
higher-order polynomial function and varies as 1/$h^{12}$, where $h$
is the distance between the interacting bodies and is difficult to
quantify.  Considering two planar surfaces, this repulsion yields an
interaction energy varying as
\begin{equation} \label{steric_general}
\phi(h)_{steric}= \frac{C}{h^8} \; ,
\end{equation}
where $C$ is a constant characterizing the interaction strength\footnote{This definition of $\phi(h)_{steric}$ here is such that
$C$ is positive and thus the interaction always repulsive.}. The reason
for the lower exponent in the flat film geometry lies in fact that
mutual interaction between all involved atoms or molecules have to
be considered. For an extended derivation of the equations we like
to refer to J. Israelachvili's approach\footnote{Here, chapter 10 in the textbook of J.
Israelachvili \cite{Israelachvili} is very helpful.}.

\textbf{Van der Waals interactions} characterize attractive
intermolecular forces of quantum-electrodynamic origin. They arise
from the variations of the zero-point energies of the collective
electromagnetic modes of the system under study. The van der Waals
energy between two molecules turns out to vary as 1/$h^6$ in the
non-retarded approximation. This approximation holds as long as the
lateral dimensions of the system are much smaller than the
wavelength of the electromagnetic fields at the dominant excitation
energies. Considering retardation effects, the interaction falls off as 1/$h^7$. In the following, we use only non-retarded potentials since in the experimental systems, effects due to retardation are mostly camouflaged by experimental error bars. Nevertheless, in some special cases including cryogenic systems, retardation effects have been reported \cite{SabiskyAnderson} and a recent theoretical study \cite{Zhao2005} discusses their possible influence on thin films in the context of the full theory of Dzyaloshinskii, Lifshitz, and Pitaevskii \cite{Dzyaloshinskii61}.

Considering again two planar surfaces, the non-retarded interaction
yields \cite{Israelachvili,Adamson}
\begin{equation} \label{vdW}
\phi(h)_{vdW}= - \frac{A}{12 \pi h^2} \; ,
\end{equation}
where $A$ is the Hamaker constant\cite{Hamaker37} and $\phi(h)$ is the
energy per unit area. A retarded interaction is described as $\phi(h)\sim A/h^3$.

The van der Waals forces of two media interacting through vacuum are
always attractive, and $A$ is positive. If the vacuum yet is
replaced by a third medium (e.g. a liquid film), things get more
involved: In a system consisting of three media $m_1/m_3/m_2$, it
may happen that $m_1$ attracts $m_2$ stronger than $m_3$ attracts
$m_2$. Thus $m_2$ is 'repelled' by $m_3$, this means that the
Hamaker constant is negative. To describe the entire system, the
Lifshitz theory\cite{Lifshitz56} has to be applied. Here, the
interacting bodies are treated as continuous media and the atomic
structure is ignored
\cite{Lifshitz56,Israelachvili,Dzyaloshinskii61}. Rather, bulk
properties as the dielectric constants and the refractive indices
are used to calculate the Hamaker constant.

For the sake of completeness, we cite here the formula
\footnote{Confer chapter 11, Eq. 11.13 of Ref.\cite{Israelachvili}}
for the Hamaker constants from the book of J.
Israelachvili\cite{Israelachvili}, which has proven to be very
useful. It is valid for two media 1 and 2, interacting through media
3. All media are taken as being dielectric with a single electronic
absorption frequency $\nu_e$, which is typically in the range of
$3\cdot 10^{15}$ Hz; $n$ is the refractive index of the medium in
the visible and $\epsilon$ the dielectric constant (i.e. $n_i^2=
\epsilon_i$, taken in the visible spectral range)
\begin{eqnarray}
A &=&A_{\nu=0} + A_{\nu>0} \\
&\approx & \frac{3}{4}kT \left( \frac{\epsilon_1-\epsilon_3}{\epsilon_1+\epsilon_3} \right) \left( \frac{\epsilon_2-\epsilon_3}{\epsilon_2+\epsilon_3} \right)\\
&& + \frac{3h\nu_e}{8 \sqrt{2}} \frac{(n_1^2-n_3^2)(n_2^2-n_3^2)}{\sqrt{(n_1^2+n_3^2)}\sqrt{(n_2^2+n_3^2)}\{\sqrt{(n_1^2+n_3^2)}+\sqrt{(n_2^2+n_3^2)}\}}
\label{Israelachvili_Hamaker1}
\end{eqnarray}

\vspace{0.3cm} The Hamaker constant $A$ will then give the strength
of the van der Waals forces between the two interfaces solid/liquid
and liquid/air.

For air as medium 1, Si as medium 2 and polystyrene as medium 3,
Eq.~\ref{Israelachvili_Hamaker1} gives\footnote{The error of the
Hamaker constant is a result of the uncertainties of the optical
properties of the involved materials as found in literature. The
error for the last digit is given in brackets.} $A_{Si}= -
2.2(5)\cdot10^{-19}$~J, replacing Si as medium 2 by silicon dioxide,
the Hamaker constant is $A_{SiOx}=1.8(4)\cdot10^{-20}$~J. For a
silane (OTS) covered Si wafer, the Hamaker constant for the
interaction with PS is calculated to be
$A_{OTS}=1.9(3)\cdot10^{-20}$~J.  For $A < 0$ ($A > 0$), the system
can gain energy by enlarging (reducing) the distance $h$ between the
surfaces, in other words, a polystyrene film of thickness $h$ is
stable on Si, since $A < 0$.

\begin{figure}[h!]
\begin{center}
\includegraphics[width=0.6\textwidth]{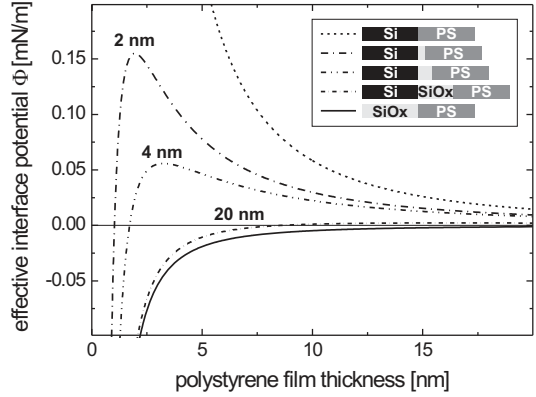}
\end{center}
\caption{\footnotesize Long-range part of the effective interface
potential $\phi(h)$ as function of PS film thickness $h$ for
different SiOx layer thickness ranging from 0 nm (dotted line) to infinity (solid line), calculated with the formula given in Eq.~(\ref{stratified}). The Hamaker constants were calculated from the optical constants of the involved materials\cite{Israelachvili}.} \label{lr-part}
\end{figure}

In Fig.~\ref{lr-part}, the dotted line represents the van der Waals
potential, $\phi(h)_{vdW}$, as given by Eq.~\ref{vdW} for a
polystyrene film on a Si wafer, where $A_{Si}$ is negative. The
potential therefore is positive, purely repulsive \footnote{The
terms 'attractive' and 'repulsive' can be misleading if thinking of
the thin film, yet the terms are chosen for a system of two media
(in our case solid and air) interacting through a third one, here
the liquid layer. Hence attractive and repulsive are meant for air
being attracted to the solid surface or 'repelled'. In case it is
repelled, the effective interface potential is repulsive for the
respective thickness of the liquid layer and the liquid wets the
solid.} and the PS film will be stable. However, $\phi(h)_{vdW}$ for
PS on an infinitely thick SiOx layer (solid line) is always
negative, purely attractive and the PS film will be unstable, since
the Hamaker constant $A_{SiOx}$ is positive.

Due to the 1/$h^2$ dependence of the potential, the van der Waals
forces are long-range forces and act significantly on distances up
to about 100 nm. In stratified systems with more than one layer
between the two half spaces of media 1 and 2, all mutual
interactions have to be considered. The exact calculation of the van
der Waals potential may therefore be quite cumbersome. This is an
important point, since there is a number of publications that
disregard this aspect and claim that theory and experiment do not
match. A system like a PS film in air, spun on a thin silicon oxide
layer on top of a Si wafer involves two thin layers (three
interfaces, abbreviated as air/PS/SiOx/Si) and can be described by
two Hamaker constants. So far, no discrepancies between theory and
experiment have ever been observed when the formulas given above had
been used. One on the most impressive successes is the quite
accurate prediction of the critical wetting temperature of pentane
on water \cite{RagilBonn}.

Assuming additivity of forces, the system air/PS/SiOx/Si can be
characterized by a summation of the van der Waals contributions of each of
the single layers with thickness $h$ of the PS and $d_{SiOx}$ of the SiOx layer:
\begin{equation}
\phi_{vdW}(h) = - \frac{A_{SiOx}}{12 \pi h^2} + \frac{A_{SiOx}- A_{Si}}{12
\pi (h + d_{SiOx})^2}  \; ,
\label{stratified}
\end{equation}
where $A_{SiOx}$ and $A_{Si}$ are the Hamaker constants of the
respective system air/PS/SiOx and air/PS/Si. Although not exact,\footnote{It should noted here that numerous studies have tried to
use only one so-called 'effective Hamaker constant' for stratified
systems using combining rules of the form
$A_{132} \approx (\sqrt{A_{22}} - \sqrt{A_{11}}) (\sqrt{A_{22}} - \sqrt{A_{33}})$. However, these rules can only achieve good
results under severe limitations, e.g., if the zero-frequency
contribution $A_{\nu=0}$ in Eq.~(\ref{Israelachvili_Hamaker1}) is
negligible \cite{Israelachvili}. Strictly speaking, they are plain
wrong, and should be used, if at all, with utmost care. It is not
surprising that many of the studies which have used them could not
quantitatively reconcile the theoretical description with the
experimental results.} the concept allows to calculate the interaction energies even for
stratified systems. With the help of Eqs.~(\ref{Israelachvili_Hamaker1}) and
(\ref{stratified}), the van der Waals potential of the
experimental system is accessible, if the SiOx layer thickness is
known and the Hamaker constants are calculated as shown before.
Fig.~\ref{lr-part} depicts the van der Waals potential as gained
from Eq.~(\ref{stratified}). Clearly, the potential is influenced by
the thickness of the silicon dioxide layer.

So the widely used system air/PS/OTS/SiOx/Si can be described by
\begin{equation}
\phi_{vdW}(h) = - \frac{A_{OTS}}{12 \pi h^2} + \frac{A_{OTS}- A_{SiOx}}{12
\pi (h + d_{OTS})^2} + \frac{A_{SiOx}- A_{Si}}{12
\pi (h + d_{SiOx} + d_{OTS})^2} \; ,
\label{stratified_OTS}
\end{equation}
yet the long-range potential $\phi_{vdW}$ is not much different to a
system with an SiOx layer of thickness $d_{OTS}+ d_{SiOx}$, since
the optical properties of the OTS and the SiOx layer are very
similar. The difference in wetting behavior will only be obvious if
the short-range potential is added.

\vspace{0.3cm}

To conclude so far, $\phi(h)$ combines short- \textbf{and}
long-range interactions. Short-range interactions are difficult to
quantify, yet van der Waals interactions can be captured by the
optical properties of the involved materials, even for stratified
systems, where the suitable van der Waals potential must be taken.
If $\phi''(h)< 0$, spinodal dewetting is possible and a
characteristic wavelength $\lambda_{s}$ should be imprinted on the
dewetting pattern.

\section{Experiments: Linking the effective interface potential to macroscopic properties}

\begin{figure}[h!]
\begin{center}
\includegraphics[width=1\textwidth]{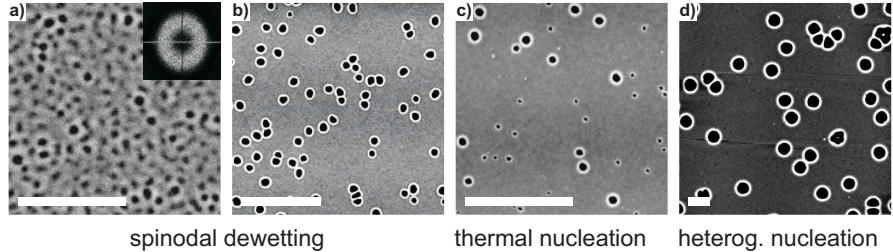}
\end{center}
\caption{\footnotesize a)-d) AFM images of dewetting PS(2k) films on Si wafers with variable Si dioxide layer and PS film thickness.
Scale bars indicate 5~$\mu$m, z-scale ranges from 0 (black) to 20~nm
(white): a) 3.9~nm PS film on a Si wafer with $d_{SiOx}=$191~nm. The inset shows a Fourier transform of the image. b) 3.9(2)~nm PS,  c) 4.1~nm PS and  d) 6.6~nm PS on Si wafers with $d_{SiOx}=$2.4~nm. (The statistical analysis of the distribution of hole sites in cases (b) to (d) was performed on larger sample areas.)}\label{dewetting_patterns}
\end{figure}

In this section experimental results shall be compared to the
predictions made by the effective interface potential. Typical
experimental dewetting patterns in the system air/PS/SiOx/Si are
shown in Fig.~\ref{dewetting_patterns}. What can we learn from
analyzing the dewetting films? I) By AFM, the contact angle $\theta$
of droplets in the late stage of dewetting can be measured. II) The
equilibrium film thickness $h_{eq}$ can be determined, e.g. by
ellipsometry. III) The spinodal wavelength $\lambda$ and IV) the
growth time $\tau$ of the spinodal pattern can be determined. V) The
morphology of the pattern. VI) The growth of the size of the holes
and VII) the form of the liquid front as function of time.

\begin{figure}[h!]
\begin{center}
\includegraphics[width=0.9\textwidth]{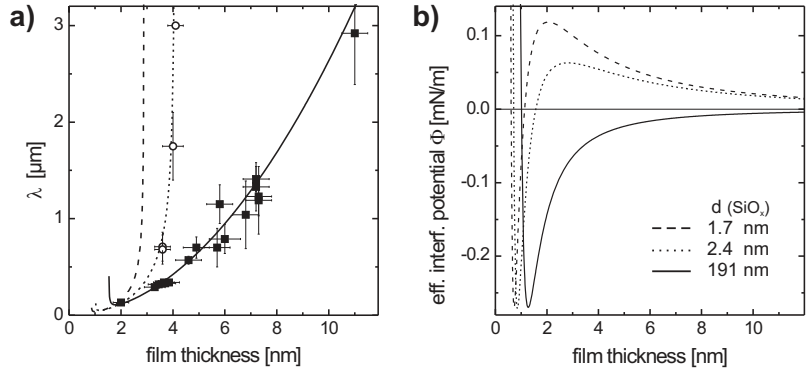}
\end{center}
\caption{\footnotesize a) Spinodal wavelength as function of PS film
thickness on Si wafers with $d_{SiOx}$ = 2.4 nm (open symbols) and
with $d_{SiOx}$ = 191 nm (filled symbols). b) The effective interface
potential $\phi(h)$ for three different SiOx layer thicknesses.} \label{Lambda_Grenzpot}
\end{figure}

Dewetting patterns like the one in Fig.~\ref{dewetting_patterns}a
clearly exhibit the spinodal wavelength. Fig.~\ref{Lambda_Grenzpot}a
comprises experiments with different PS film thicknesses on Si
wafers with $d_{SiOx}$ = 2.4 nm (open symbols) and with $d_{SiOx}$ =
191 nm (filled symbols). From the data of $\lambda(h)$ data points
for $\phi''(h)$ can be gained. Now, fitting the second derivative of
Eq.~(\ref{stratified}) to the data, the Hamaker constants and the
short-range interaction constants can be obtained as fit parameters
and the entire effective interface potential can be inferred. It is
plotted in Fig.~\ref{Lambda_Grenzpot}b.

The best fits\cite{See2001,Seemann2001} are achieved for
$A_{Si,fit}= - 1.3(6)\cdot10^{-19}$~J and
$A_{SiOx,fit}=2.2(4)\cdot10^{-20}$~J. The values for $A$ nicely
match the values calculated from optical properties of the media
involved, as described by Eq. \ref{Israelachvili_Hamaker1}. For C we
find  $C_{SiOx,fit}= 6.3(1)\cdot10^{-76}$~Jm$^6$ and $C_{OTS,fit}=
2.1(1)\cdot10^{-81}$~Jm$^6$. In the cited references, a detailed
description of the fitting and reconstruction procedure can be
found.

\begin{figure}[h!]
\begin{center}
\includegraphics[width=0.9\textwidth]{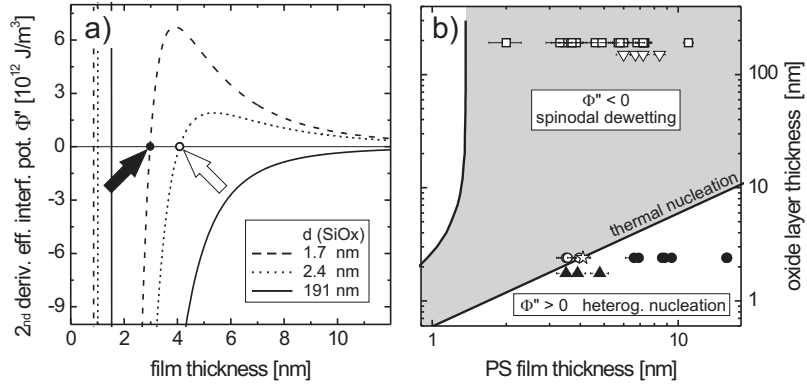}
\end{center}
\caption{\footnotesize a) Second derivative of the effective
interface potential as function of film thickness for three
different SiOx layer thicknesses as fitted to the data of
Fig.~\ref{Lambda_Grenzpot}. b)Stability diagram of PS films on top
of Si wafers with variable oxide layer
thickness.\cite{See2001}}\label{secondderiv_stability}
\end{figure}

The second derivative $\phi''(h)$ is plotted in in
Fig.~\ref{secondderiv_stability}a for three SiOx layer thicknesses.
An arrow marks the zero in the respective second derivative,
corresponding to the relevant inflection points of the curves
displayed in Fig.~\ref{Lambda_Grenzpot}b. Spinodal dewetting is
possible only for films thinner than indicated by the arrows: PS
films on Si wafers with an 1.7~nm (2.4~nm) thick oxide layer are
unstable for PS film thicknesses below 3~nm (4~nm). For larger
thicknesses, the films are metastable and need to overcome a potential
barrier in order to finally reach the global minimum of $\phi$. The
maximum in $\phi$, which is visible in Fig.~\ref{Lambda_Grenzpot}b
for the broken curves, represents only part of this barrier, since a
nucleus for dewetting as it is formed, e.g., by thermal activation,
is a localized structure and thus involves excess surface energy as
well \cite{Blossey95}. It is readily checked that the energy
required to form such nucleus is in almost all cases large as
compared to $kT$, such that thermal nucleation plays no role in
systems of practical interest. This can be illustrated nicely with
the PS films, as shown in Fig.~\ref{secondderiv_stability}b. PS
films on thick 191 nm SiOx (solid line in
Fig.~\ref{Lambda_Grenzpot}a) are unstable for all relevant film
thicknesses.
%(For films thicker than 10 nm, the instability just may take too long to be observable.)
On 2.4 nm of oxide, however, a sign reversal is observed at a film
thickness of about 4 nm (white arrow in Fig.~\ref{secondderiv_stability}a). Only very close to this point, the height of the
potential barrier vanishes, and homogeneous nucleation by thermal
activation is possible \cite{Blossey95,HerminghausBrochard,TsuiThermal}.
However, nucleation can as well, and usually does, proceed by means
of localized defects in the film. A defect, be it in the molecular
texture of a polymeric film or just a small dust particle, may
remove the potential barrier locally and thus induce dewetting. This
rupture mechanism is termed `heterogeneous nucleation'
\cite{Mitlin93,Mitlin94}, in analogy to defect-mediated nucleation
in demixing scenarios.

The fact that the effective interface potential can be directly
inferred from macroscopic quantities, such as the spinodal wavelength, is one
reason why experimentalists are seeking for spinodally dewetting
regimes in thin liquid film systems. However, if a company is asking why their coating does not
stay stable on a surface, it is convenient to rather have a stability
diagram which allows to look up where the system will be in a metastable or an
unstable state. A stability diagram for the system air/PS/SiOx/Si
is shown in Fig.~\ref{secondderiv_stability}b. The solid lines of
the diagram base on Eq.~\ref{effective_interface} with a long-range
potential as in Eq.~\ref{stratified} and separate the spinodal
dewetting (unstable) regime, where $\phi''(h)< 0$, from in the
regime of heterogeneous nucleation, characterized by $\phi''(h) > 0$
(metastable regime). Thermal nucleation is possible for $\phi''(h) =
0$, this line separates the two regimes. Experiments that exhibit
spinodal dewetting patterns are indicated by open symbols, those of
heterogeneous nucleation (randomly distributed holes) are indicated
by solid symbols\footnote{As explained above, thermal nucleation can be only observed if
the nucleation barrier is of order $kT$. On the scale of
Fig.~\ref{secondderiv_stability}b, this is fulfilled only in a
region smaller than the width of the line denoted 'thermal
nucleation'. Alleged observations of thermal nucleation in a wider
range \cite{DeSilva07} appear very questionable.}. A star marks the
set of parameters where thermal nucleation was observed. Note that in
the unstable regime, heterogeneous nucleation from localized defects is also possible and
indeed is sometimes observed. In fact, it usually dominates if the experiments are not performed in extremely clean conditions.
Spinodal dewetting, however, can only take place in the unstable regime.

We can state here that the effective interface potential indeed is
the key for wetting properties of liquids on surfaces. It can easily
be obtained via the optical properties of the involved media and by
taking into account the adapted van der Waals potential for
stratified systems. It is worth noting that metastable
potentials can only be present if there is more than one Hamaker
constant involved, one with negative and one with positive sign.

Coming back to the company - a stability diagram for their specific
system will reveal the way to a solution of their wettability
problem: If their coating system is located in the metastable regime
(in the example shown, it depends on the silicon dioxide thickness
and the polystyrene thickness), the advice is to reduce the number
of possible nucleation centers, e.g. by creating a cleaner
environment. If the coating system yet is in the unstable regime,
the introduction of an adhesion promoter is necessary.

As mentioned at the beginning, the shape of a droplet close to the
three phase region is influenced by the long-range intermolecular
interactions. By AFM, this region can be characterized. For a
detailed discussion we like to refer to the work of T. Pompe et
al.\cite{Pompe00, Pompe02}. In Fig.~\ref{drop-profile} two profiles
are sketched that were calculated based on the knowledge of the
effective interface potential. Fig.~\ref{drop-profile}a is a
characteristic curve for an unstable potential as shown for
$d_{SiOx}= 191~nm$ in Fig.~\ref{Lambda_Grenzpot}a, whereas a profile
like Fig.~\ref{drop-profile}b is expected for a metastable potential
like for $d_{SiOx}= 1.7~nm$. The experimental AFM results (not shown
here) corroborate the theoretical expectations\cite{Seemann05}.

One question is open up to now: are spinodal dewetting patterns
always as easy to detect as in Fig.~\ref{dewetting_patterns}a? The
answer is 'no': patterns which at first glance appear like the one
shown in Fig.~\ref{dewetting_patterns}b can be generated due to a
spinodal process as well \cite{Herminghaus98}. The next section we
will briefly explain how to distinguish a dewetting pattern of a
system in the metastable state from one in the unstable regime and
also address thermal nucleation.
\vspace{1.3cm}

\section{Characterizing experimental dewetting patterns}

The three rupture scenarios \textbf{spinodal dewetting,
homogeneous nucleation}, and \textbf{heterogeneous nucleation} give rise to specific
dewetting patterns. Vice versa, characterizing the experimental
dewetting pattern can help to identify the rupture mechanism and to
infer the effective interface potential. Fig.~\ref{PS-Film} and
Fig.~\ref{dewetting_patterns} show experimental examples of films
dewetting via different rupture mechanisms. Theoretically, the distinction between nucleation and spinodal
dewetting appears quite clear: Vrij \cite{Vrij} proposed already in 1966
that a \textbf{spinodal rupture} of a free liquid film results in a
dewetting pattern of `hills and gullies' with a preferred distance
$\lambda_s$ after a certain time of rupture $\tau$.

Experimentally, the rupture time $\tau$ is difficult to measure since the hole which forms as the film has thinned to
zero thickness must
have a certain size to be observable. Experimentalists thus
concentrated instead on the evidence of a preferred wavelength $\lambda_s$
observable in their systems.\footnote{Note that a preferred hole distance
$\lambda_s$ has to be distinguished from a mean hole distance:  In
some studies it was found that the mean hole distance scaled with
the film thickness as expected for spinodal dewetting
\cite{Reiter92,Reiter93}. Therefore, the scenario shown in
Fig.~\ref{PS-Film}a was often regarded as a typical spinodal
dewetting scenario. However, it turned out later that this was not
correct:\cite{Jacobs98} first of all, the system is metastable for
the film thicknesses studied, as an analysis similar to
Eq.~\ref{stratified} would have shown. Secondly, the typical time
scale for spinodal dewetting does not fit the theoretical
expectation (rupture time $\tau \propto h^5$).} If, however, the holes are randomly (Poisson) distributed, they are
assumed to stem from \textbf{heterogeneous nucleation}, reflecting
the fact that defects typically exhibit random statistics. Although
it was generally accepted later that (heterogeneous) nucleation from localized defects is the
reason for the dewetting scenario shown in Fig.~\ref{PS-Film}a, the
very nature of the nucleation defect mostly remains unclear: it might be
dust particles or any other chemical or physical inhomogeneities. In
some holes, by light microscopy or atomic force microscopy (AFM), a
nanoscopic object could be observed right in the center of each
hole. Assuming the object to be a dust particle, the first trial was
to reduce the number of nucleation centers by improving the preparation
conditions. The number of holes, however, could not be reduced below
a certain level, which suggests that the physics of hole nucleation
in polymer films may be deeper than a mere effect of `dirt'. It has
been shown experimentally that stress inside the thin films
(stretched entanglements\cite{Croll}) caused by the preparation of the film out
of solution plays a significant role and can cause
holes. Details can be found in the studies of Seemann et
al.\cite{Seemann05} and Reiter et al. \cite{ReiterNat2005} .

\vspace{0.4cm}
The experimental distinction between a hole pattern with preferred hole
distance and a pattern with randomly distributed hole sites is far from being obvious.
The thicker the films are, the weaker is the driving force, and the longer
is the growth time $\tau$ of the spinodal mode\cite{Vrij,Ruckenstein,Seemann2001}
(typically, $\tau \propto h^5$)  and can easily exceed experimental time scales.
For thicker films, dewetting by heterogeneous nucleation may therefore be
quicker and can suppress a spinodal pattern \cite{Konnur,Neto03}.
Moreover, chemical hetero\-gene\-it\-ies can locally cause a change in
$\phi$ and therefore the rupture conditions of the sample may vary
from spot to spot leading to a less-ordered dewetting pattern. This
effect is more pronounced in thicker films due to the small driving
forces and the large growth time $\tau$. Hence a two-point correlation function
(a radial pair correlation function $g(r)$ or a Fourier transform) might not be
sensitive enough to detect the correlations between the hole sites, or the
statistics is too poor. Then, more powerful tools have to be applied. Minkowski
functionals - based on integral
geometrical methods - have shown to be a versatile method to track
down higher order correlations. There is no room here to dwell on this method in
any reasonable detail. The reader is referred to the pertinent literature
\cite{MeckeDiss,Herminghaus98,Jacobs98,JacobsMecke,Becker03,Renate2007b}.

Let us come back to thermal nucleation once more. As discussed above,
the potential barrier for nucleation of a dry spot may in principle be overcome by thermal activation,
provided the system is sufficiently close to the sign reversal of $\phi''(h)$.
The characteristic feature of this scenario is a continuous breakup of more and more holes, whereas
nucleation from defects causes holes that emerge only within a sharp time window \cite{JacobsSchatz98}.
Fig.~\ref{dewetting_patterns}c depicts an example for a 4.1~nm thick PS films on a wafer
with a 2.4~nm SiOx layer: Holes of different sizes are observable. For that system, the
stability diagram of Fig.~\ref{secondderiv_stability}\,b reads that  $\phi''(4.1~nm)\approx 0$,
which is another strong indication that the theoretical predictions corroborate the experimental
observations and are able to capture the wettability of dielectric systems.

\begin{figure}[h!]
\begin{center}
\includegraphics[width=0.9\textwidth]{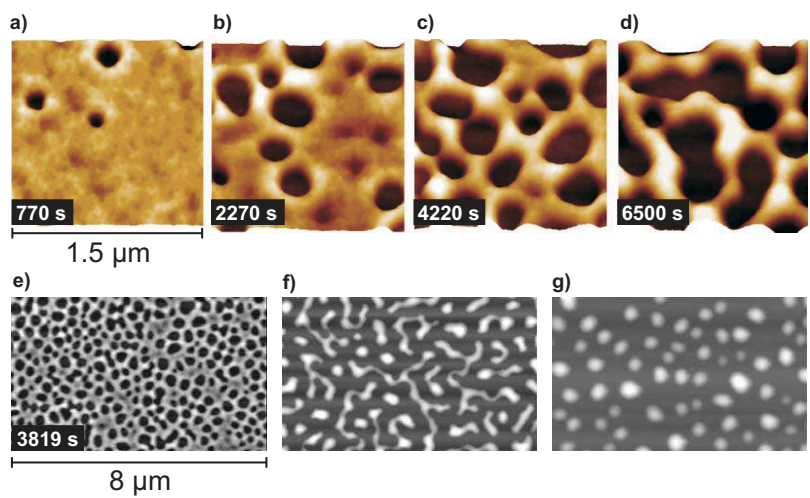}
\end{center}
\caption{\footnotesize Dewetting morphology of a 3.9(2) nm PS(2k) film on a Si wafer with a 191~nm thick SiOx layer as recorded by in situ AFM. Up to about 5000 s (e) the temperature was held constant at 53$^{\circ}$ and annealing times are given in the pictures. Afterwards, the temperature was successively increased to 100$^{\circ}$. The scan size in (a)­(d) is 1.5~ µm. In (e)­(g) larger scans were made to check possible damage of the sample by the AFM tip. Scan (e) was taken some minutes before scan (d), whereas scan (f) was recorded some minutes after scan (d). Scan (g) characterizes the end of the dewetting process and Ostwald ripening of the droplets slowly sets in.}\label{AFM_serie}
\end{figure}

\section{Dynamics of spinodal dewetting}
At the beginning of spinodal dewetting studies, only one system was known to dewet spinodally: thin gold films on top of quartz substrates \cite{Bischof}. Metal films, however, are not so easy to deal with as compared to polymer films, since the films need to be annealed by a laser. Hence, the time scale is in the ns range, crystallization plays a role and the Hamaker constants are not so easy to calculate. Polymer films such as polystyrene (PS), however, are dielectric and Hamaker constants can easily be determined via Eq.~\ref{Israelachvili_Hamaker1}.

\begin{figure}[h!]
\begin{center}
\includegraphics[width=0.6\textwidth]{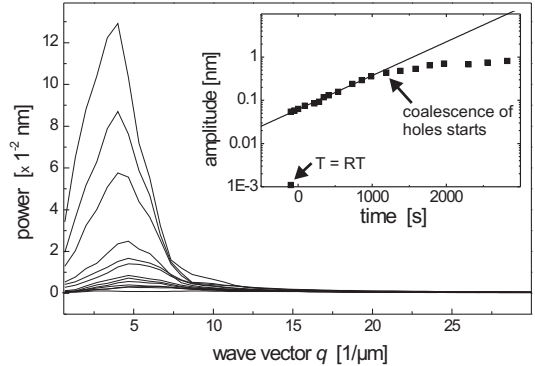}
\end{center}
\caption{\footnotesize Results of Fourier transforms giving the power spectral density of in situ AFM pictures at $T = 53^{\circ}$C, on of which is shown in Fig.~\ref{dewetting_patterns}a. Note that the time intervals between the curves are not constant. The inset depicts the amplitude of the undulation as a function of annealing time. The first data point at $t < 0$ s gives the roughness of the PS film surface at room temperature (RT) as revealed from a Fourier transform. The solid line is a fit of an exponential growth to the data, as expected from theory\cite{Ruckenstein}.}\label{exp-Anwachsen}
\end{figure}

In situ AFM studies on a dewetting PS film on a Si wafer with a thick SiOx layer reveal the dynamics of the structure formation process, c.f. Fig.~\ref{AFM_serie}. Fig.~\ref{exp-Anwachsen} depicts results of Fourier transforms of the AFM scans.  Clearly, one mode is amplified fastest. The amplitude of the growing unstable mode, shown in the inset, grows exponentially with time, as expected from theory. Further investigations show that at the onset of the dewetting, no mode is specifically selected, rather, the amplitude reflects the roughness of the film. Comparing the results to deterministic simulations,\cite{Becker03} the pattern formation process and the morphology of the dewetting structures match very well, yet the temporal evolution of the morphology slightly differs. It turned out that thermal fluctuations accelerate the dewetting dynamics in the experiments.\cite{TsuiThermal,Renate2007b}

\section{Conclusion}
The wettability of a substrate is a delicate interplay of forces. For dielectric systems, the wettability can be described successfully by the effective interface potential $\phi$, which is a sum of short- and long-range interactions. The latter are dominated by van der Waals interactions. Their strength can be obtained by calculating the Hamaker constant via the optical properties of the involved media. For stratified systems, the additivity of forces can be assumed.

Rupture mechanism, pattern formation, morphology and dynamics of dewetting are all governed by the effective interface potential and the experiments can corroborate the theoretical expectations.

The knowledge of the effective interface potential therefore allows for a tailoring of the wettability on demand and is a great tool for further fundamental or applied studies.

%\begin{appendix}[Optional Appendix Title]
%\section{Sample Appendix}
%Text...
%\end{appendix}

%\bibliographystyle{ws-rv-van}
%\bibliography{ws-rv-sample}

%\printindex[aindx]                % to print author index
\printindex

\end{document}